\documentclass[aps,prd,floatfix,onecolumn,reprint,amsmath,amssymb,superscriptaddress]{revtex4-2}
\usepackage{amsfonts}
\usepackage{mathrsfs}
\usepackage{amsmath}
\usepackage{color}
\usepackage{natbib}
\usepackage{graphicx}
\usepackage{bm}
\usepackage{amssymb}
\usepackage{mathptmx}
\usepackage{stmaryrd}
\usepackage{xspace}
\usepackage{epstopdf}
\usepackage{dcolumn}
\usepackage{longtable}
\usepackage{multirow}
\usepackage{bbding}
\usepackage{amsthm}
\usepackage{subfiles}
\usepackage{blindtext}
\usepackage{gensymb}

\usepackage[colorlinks=true, letterpaper=true, pdfstartview=FitV, linkcolor=blue, citecolor=blue, urlcolor=blue]{hyperref}

\makeatother

\begin{document}
\title{Half-Metallic Ferromagnetic Weyl Fermions Related to Dynamic Correlations in the Zinc-blende Compound VAs} 

\author{Xianyong Ding}
\altaffiliation{X. Y. D. and X. J. contributed equally to this work.}
\affiliation{Institute for Structure and Function $\&$ Department of Physics, Chongqing University, Chongqing 400044, China}
\affiliation{Center for Quantum materials and devices, Chongqing University, Chongqing 400044, China}

\author{Xin Jin}
\altaffiliation{X. Y. D. and X. J. contributed equally to this work.}
\affiliation{College of Materials Science and Engineering, Chongqing Normal University, Chongqing 400044, China}

\author{Haoran Wei}
\affiliation{Institute for Structure and Function $\&$ Department of Physics, Chongqing University, Chongqing 400044, China}
\affiliation{Center for Quantum materials and devices, Chongqing University, Chongqing 400044, China}

\author{Ruixiang Zhu}
\affiliation{Institute for Structure and Function $\&$ Department of Physics, Chongqing University, Chongqing 400044, China}
\affiliation{Center for Quantum materials and devices, Chongqing University, Chongqing 400044, China}

\author{Xiaoliang Xiao}
\affiliation{Institute for Structure and Function $\&$ Department of Physics, Chongqing University, Chongqing 400044, China}
\affiliation{Center for Quantum materials and devices, Chongqing University, Chongqing 400044, China}

\author{Xiaozhi Wu}
\affiliation{Institute for Structure and Function $\&$ Department of Physics, Chongqing University, Chongqing 400044, China}
\affiliation{Center for Quantum materials and devices, Chongqing University, Chongqing 400044, China}

\author{Fangyang Zhan}
\email{zhan_fyang@cqu.edu.cn}
\affiliation{Institute for Structure and Function $\&$ Department of Physics, Chongqing University, Chongqing 400044, China}
\affiliation{Center for Quantum materials and devices, Chongqing University, Chongqing 400044, China}

\author{Rui Wang}
\email{rcwang@cqu.edu.cn}
\affiliation{Institute for Structure and Function $\&$ Department of Physics, Chongqing University, Chongqing 400044, China}
\affiliation{Center for Quantum materials and devices, Chongqing University, Chongqing 400044, China}

\begin{abstract}
The realization of 100\% polarized topological Weyl fermions in half-metallic ferromagnets is of particular importance for fundamental research and spintronic applications. Here, we theoretically investigate the electronic and topological properties of the zinc-blende compound VAs, which was deemed as a half-metallic ferromagnet related to dynamic correlations. Based on the combination of density functional theory and dynamical mean field theory, we uncover that the half-metallic ferromagnet VAs exhibit attractive Weyl semimetallic behaviors with twelve pairs of Weyl points, which are very close to the Fermi level. Meanwhile, we also investigate the magnetization-dependent topological properties; the results show that the change of magnetization directions only slightly affects the positions of Weyl points, which is attributed to the weak spin-orbital coupling effects. The topological surface states of VAs projected on semi-infinite (001) and (111) surfaces are investigated. The Fermi arcs of all Weyl points are clearly visible on the projected Fermi surfaces. Our findings suggest that VAs is a fully spin-polarized Weyl semimetal with many-body correlated effects for spintronic applications.
\end{abstract}



\maketitle

Half metallic ferromagnets (HMFs) which are metals for one spin polarization and semiconductors for the other, were first reported in XMnSb (X=Ni, Pt) by Groot et al. \cite{de1983new,de1984half,de1986recent}. HMFs that show the behaviors of a full spin-polarization of carriers are necessary to boost the efficiency of spintronic devices \cite{BENMAKHLOUF2018430}. It is also vital for fundamental research, particularly for the origin of half-metallic gaps, nature of interatomic exchange interactions \cite{SaSioglu_2005}, or effects of electronic correlations \cite{Irkhin_1994}. Due to their potential applications in spintronics and abundant correlation physics, the HMFs have attracted intensive attention and growing interest \cite{Irkhin_1994, RevModPhys.76.323,chappert2007emergence}. In recent decades, such electronic features are found in many magnetic materials, including perovskite structures \cite{soulen1998measuring}, CrO$_{2}$ \cite{korotin1998cro}, Fe$_{3}$O$_{4}$ \cite{yanase1984band}, Heusler alloys \cite{galanakis2002origin,galanakis2002slater,galanakis2004appearance,shreder2008evolution,fomina2011electrical,shreder2017optical,lidig2019surface,jourdan2014direct}, diluted magnetic semiconductors \cite{akai1998ferromagnetism,sandratskii2002exchange}, spin gapless semiconductors \cite{wang2008proposal,ouardi2013realization}, and topological semimetals (TSMs) \cite{manna2018heusler,wang2017dirac}, etc.

In binary transition metal compounds, two-dimensional (2D) XY (X = V, Cr; Y = P, As) and three-dimensional (3D) systems of MnAs, CrAs, and MnC are found to be HMF materials \cite{PhysRevB.104.104417,PhysRevB.102.024441,PhysRev.139.A796,PhysRev.139.A796,PhysRevB.83.113102}. The zinc-blende compounds of transition elements V, Cr, and Mn with $sp$ elements N, P, As, Sb, S, Se, and Te were reported to exhibit 100$\%$ spin polarization and half-metallic behaviors; and especially, the zinc-blende compounds VAs, VSb, CrAs, CrSb, VTe, CrSe, and CrTe were suggested to be promising candidate materials \cite{PhysRevB.67.104417}. Among these compounds, CrAs have been successfully synthesized and have been confirmed to possess half-metallic properties both by experiments and first-principles calculations \cite{Akinaga_2000, doi:10.1063/1.1558604}. Another important compound in this family materials is VAs, which was reported to be a ferromagnetic semiconductor with high Curie temperature \cite{PhysRevB.67.104417, PhysRevB.68.054417} using density functional theory (DFT) calculations but a half-metallic ferromagnetic state using  DFT plus dynamical mean field theory (DFT+DMFT) approach \cite{PhysRevLett.96.197203}. 

On the other hand, a closing of band gap induced by dynamical correlations in VAs would potentially accompany with topological transitions. It is well-known that Weyl semimetals (WSMs) \cite{PhysRevB.83.205101, PhysRevLett.107.186806} are widely studied theoretically and experimentally \cite{lv2015experimental,lu2015experimental,xu2015discovery,borisenko2019time}. Most theoretical efforts for predicting WSMs are supported by first-principles calculations within the DFT framework \cite{wan2011topological, weng2015weyl, wang2016time, nie2022magnetic}. The HMFs may be in the case of strongly correlated electron systems. For such systems, promising topological states related promising topological states would be present. Considering that the zinc-blende VAs lack spatial inversion symmetry and time-reversal symmetry, half-metallic ferromagnetic Weyl fermions with the dynamical many-electron feature are expected.

In this work, based on density functional theory and dynamical mean field theory, we investigate the correlation effect in zinc-blende compound VAs. We first calculate the electronic properties of VAs based on the DFT + DMFT method in the absence of spin-orbit coupling, in which the validations are confirmed in many strong correlation systems. By adjusting the effect $U$ values of V-d orbitals in the DFT plus $U$ approaches, the spin-polarized electron structure and density of states using the DFT + $U$ and DFT+DMFT methods are compared. We find that the electronic properties around the fermi level of DFT + $U$ methods can be well in agreement with that of considering the dynamical many-body correlation effects when the effective $U$ values are set to 1.5 eV for V-d orbitals. As the topological properties are only related to the states around the fermi level, the following calculations of this paper are under the framework of the DFT + $U$ method. The magnetic moment is 2$\mu$B for V atom per unit formula. The topological surface states and Fermi arcs of (001) and (111) surfaces of VAs can be clearly visible in different energies. Besides, there exist twelve pairs of Weyl Points (WPs) in the entire Brillouin Zone of VAs in the magnetic direction of (001) and (111). The Weyl Points (WPs) are closed to the Fermi level and distributed asymmetrically in the first Brillouin Zone, due to the absence of time-reversal and inversion symmetries. The topological charges of these Weyl Points show that it is a magnetic Weyl semimetal with HMF properties. These results indicate that VAs are important in fundamental research of correlation physics and spintronic applications.

\begin{figure}[!t]
\includegraphics[width=\linewidth]{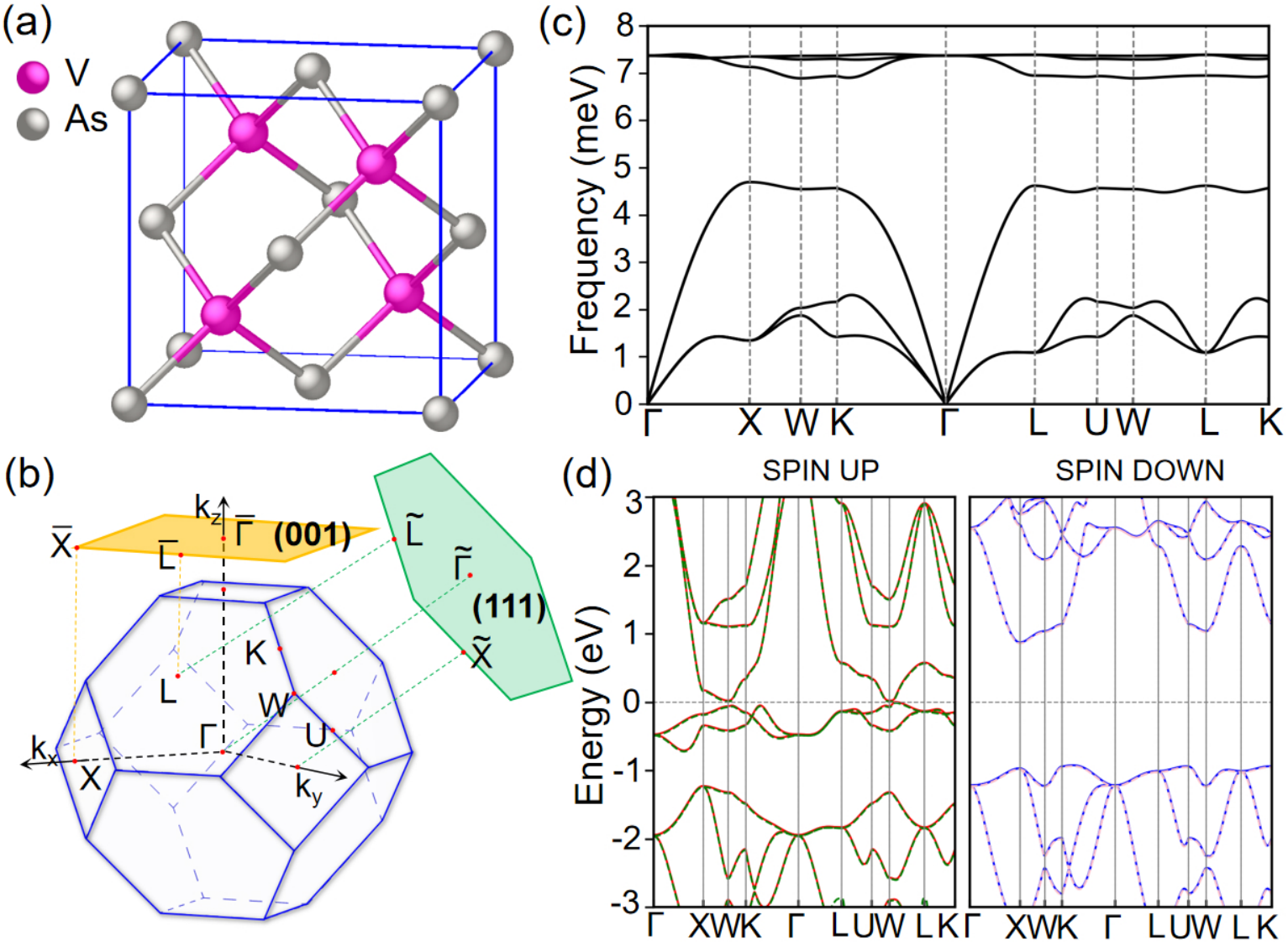}
\caption{(a) Crystal structure of the unit cell for VAs with space group of $F\overline{4}3m$ (No. 216), in which the red ball represents V element and gray ball represents As element. (b) The first Brillouin Zone (BZ) of VAs, the yellow and green color planes represent the (001) and (111) surfaces BZ. (c) Phonon spectrum of VAs. (d) The spin-polarized band structures of VAs. The red and blue solid lines denote different spin components computed with PBE functionals, and the green and pink dotted lines represent the corresponding results constructed with MLWFs.}
\label{fig1}
\end{figure}

As depicted in Fig. \ref{fig1}(a), VAs crystallize in an FCC lattice with a space group of $F\overline{4}3m$ (No. 216). The lattice parameters are optimized to 6.26 $\mathrm{\AA}$ with $U_{eff}$ = 1.5 eV which is a little different from the values of the previous result because of the different U values \cite{PhysRevLett.96.197203}. The magnetic anisotropic energies of VAs are calculated and supported in Fig. S1 of the Supplement Materials \cite{SM}. The tiny energy differences suggest the magnetic directions calculated in this work are all possible magnetic ground states of VAs. The magnetization moment of V element is obtained to 2.0 $\mu$B per unit formula of both magnetization direction [001] and magnetization direction [111]. The absence of negative phonon energies in Fig. \ref{fig1}(c) confirms the dynamical stability of the zinc-blende VAs compound. Fig. \ref{fig1}(b) illustrates the three-dimensional Brillouin Zone (BZ), high-symmetry points, and the corresponding surface BZ of (001) and (111) surfaces for VAs compound, which will be used in the following calculations. The fitting precision in Fig. \ref{fig1}(d) is obtained by using the Maximal Localized Wannier Functions (MLWFs) in the absence of the spin-orbital coupling (SOC) effect. The Wannier subspace composed by these MLWFs will be used in the following dynamical mean field calculations.

\begin{figure}[t]
\includegraphics[width=\linewidth]{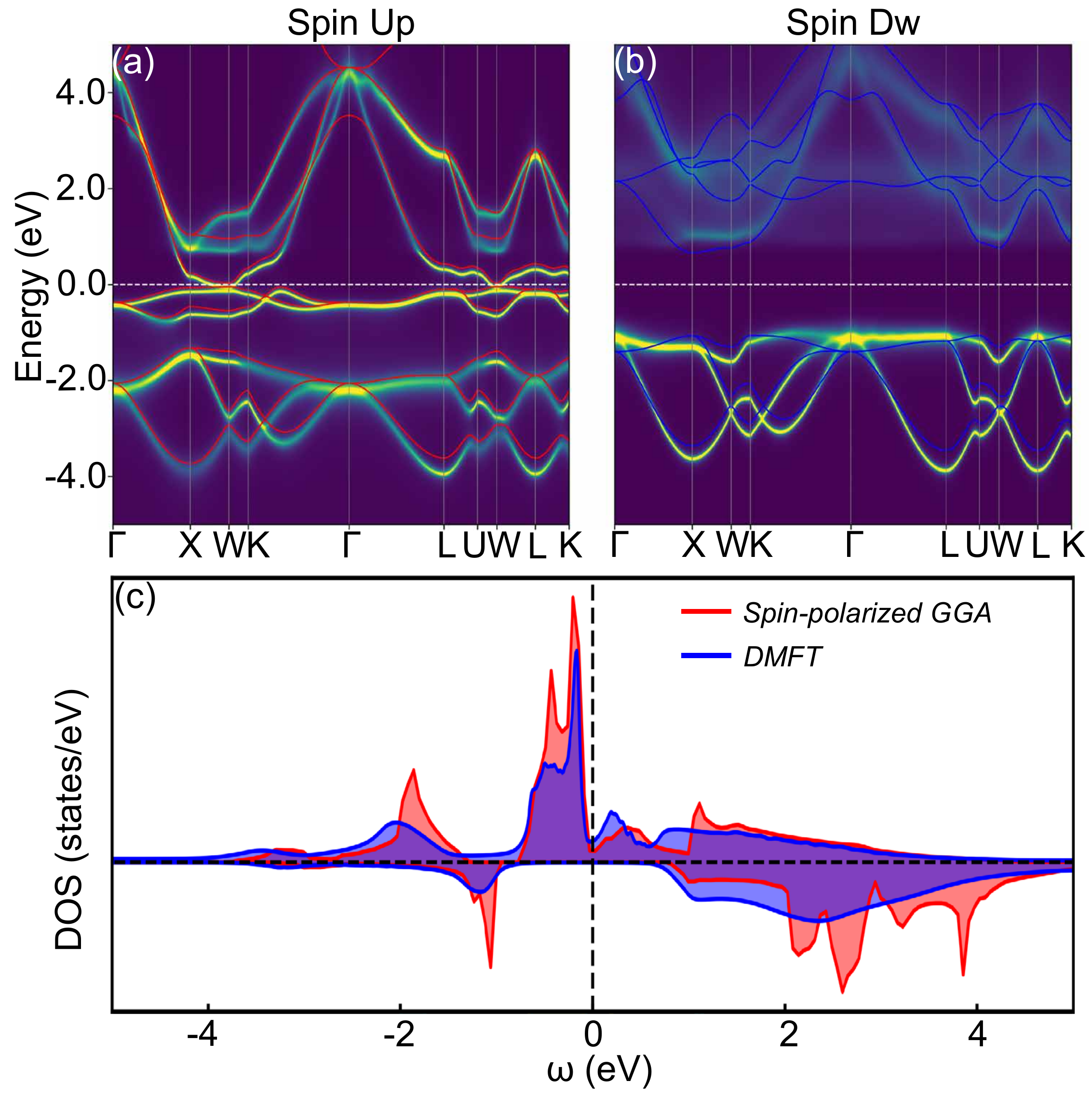}
\caption{(a, b) Band structure of VAs calculated by DFT + DMFT approach at the temperature of 300 K and DFT + $U$ method without considering SOC effect, in which the left panel represents spin-up components and the right panel represents spin-down components of VAs with V sublattice forming ferromagnetic order. (c) The density of states of V-d orbitals obtained from DFT + DMFT and DFT + $U$ methods.}
\label{fig2}
\end{figure}

It is reported in previous work \cite{PhysRevLett.96.197203}, the electronic structure of zinc-blende compound VAs shows that it is a narrow band gap semiconductor by considering the correlation effect through density functional theory LSDA + $U$ ($U$=2.0 eV, $J$=0.9 eV) method for V-d orbital. However, it shows metallic behavior when treating the correlation effect in the framework of density function theory (DFT) plus dynamical mean field theory (DMFT), i.e. DFT + DMFT method. These results reveal that the electronic properties of VAs can not be determined only by using DFT method. Here, we define V-3d orbitals as the correlation subspace to solving the DMFT equations. As displayed in Fig. \ref{fig1}(d), the MLWFs of V-3d, and As-3p are constructed by using Wannier90 \cite{RevModPhys.84.1419, MOSTOFI20142309} code over the energy range [-5.0, 4.7] eV with respect to the Fermi energy. The hybridization window is spanned by the p-d manifold, in which V-3d orbitals are treated as the correlated orbitals. Then, DFT + DMFT calculations are implemented with the correlated subspace in which the Hubbard interaction strength $U$ = 5 eV and the Hund`s interaction strength $J$ = 1 eV are used. The parameters for double-counting of $U$ as $U$ - $\alpha$ are set to DC$\_$type = 1 and $\alpha$ = 0.2. The DMFT impurity problem is solved using the continuous-time Quantum Monte Carlo (CTQMC) \cite{RevModPhys.83.349, PhysRevB.75.155113} method with a temperature of 300 K. The theoretical background of DFT + DMFT is supported in the SM \cite{SM}.

\begin{figure*}[t]
\includegraphics[width=0.8\linewidth]{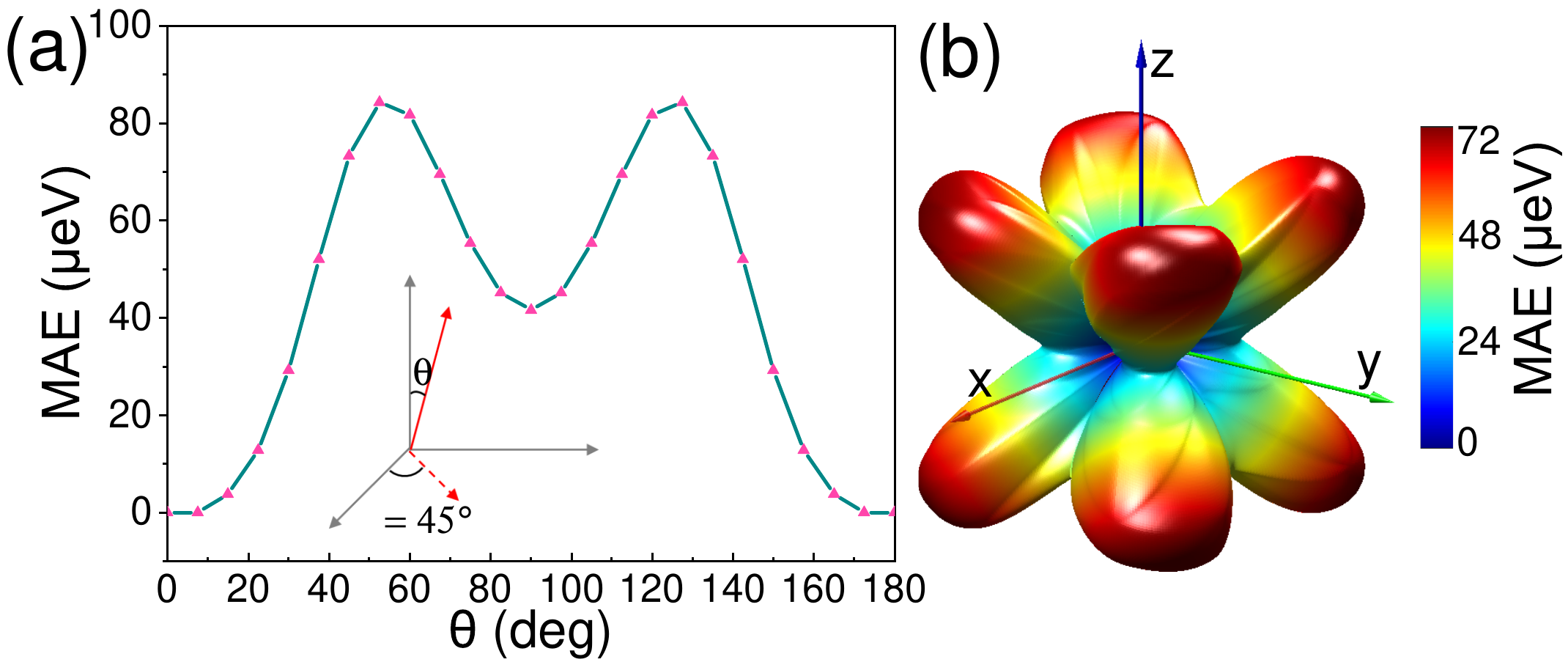}
\caption{The magnetization anisotropy energy of (a) $\phi$=45$\degree$, and (b) the energy differences of VAs in the magnetization directions of different $\theta$ and $\phi$.}
\label{fig3}
\end{figure*}

Fig. \ref{fig2}(a) and \ref{fig2}(b) are the electronic structures of VAs based on DFT + DMFT and DFT + $U$ approaches in the absence of SOC effect, in which the DFT part is calculated with the PBE exchange and correlation functional. The results clearly show that within the energy range of -5 ev to 5 eV, the band structures of the DFT + DMFT method are in good agreement with that of the DFT + $U$ ($U_{eff}$ = 1.5 eV) method. Moreover, the total density of states (TDOS) summed by V-d orbitals and As-p orbitals in Fig. S1 is consistent with previous results \cite{PhysRevLett.96.197203} suggesting the metallic properties of VAs. In Fig. \ref{fig2}(c), the DOS of V-d orbitals of DFT + DMFT is also in agreement with that of DFT + $U$ method near the Fermi level. The topological semimetal properties are only relevant to the states near the Fermi level, confirming the validation to investigate the topological semimetal properties of VAs compound by using DFT + $U$ approaches. These results reveal that DFT + $U$ method can also be a valid way to describe the correlation effect of VAs. Therefore, the electronic properties introduced in the following sections are all obtained by using the DFT + $U$ method.

\begin{table}[!b]
\caption{\label{weylpoints} The corresponding positions, topological charges, and energies of the WPs in the magnetization direction [001] of VAs.}
\renewcommand\tabcolsep{18 pt}
\begin{tabular}{ccr}      
\hline
\hline
     Coordinate (x, y, z)  & Charge   & E-E$_{F}$ (eV) \\
\hline
        (0.0, -1.0, -0.46)      &  -1    & 0.022 \\ 
        (-0.46, 0.0, -1.0)      &  -1    & -0.028 \\ 
        (0.48, -1.0, 0.0)      &   +1    & -0.023 \\ 
        (0.48, 0.0, 1.0)      &  -1    & -0.018 \\ 
        (0.46, 0.0, -1.0)      &  -1    & -0.027 \\ 
        (-0.48, 1.0, 0.0)      &   +1    & -0.023 \\ 
        (0.0, -0.47, -1.0)      &   +1    & -0.018 \\ 
        (1.0, 0.48, 0.0)      &  -1    & -0.023 \\ 
        (-0.48, -1.0, 0.0)      &   +1    & -0.023 \\ 
        (0.48, 1.0, 0.0)      &   +1    & -0.023 \\ 
        (-0.48, 0.0, 1.0)      &  -1    & -0.018 \\ 
        (1.0, 0.0, -0.50)      &   +1    & -0.073 \\ 
        (0.0, 1.0, 0.5)      &  -1    & -0.074 \\ 
        (-1.0, 0.0, -0.5)      &   +1    & -0.074 \\ 
        (0.0, -1.0, 0.5)      &  -1    & -0.073 \\ 
        (0.0, 0.45, 1.0)      &   +1    & -0.028 \\ 
        (-1.0, -0.48, 0.0)      &  -1    & -0.023 \\ 
        (-1.0, 0.48, 0.0)      &  -1    & -0.023 \\ 
        (1.0, 0.0, 0.45)      &   +1    & 0.022 \\ 
        (1.0, -0.48, 0.0)      &  -1    & -0.022 \\ 
        (0.0, 0.48, -1.0)      &   +1    & -0.018 \\ 
        (-1.0, 0.0, 0.45)      &   +1    & 0.022 \\ 
        (0.0, 1.0, -0.45)      &  -1    & 0.022 \\ 
        (0.0, -0.46, 1.0)      &   +1    & -0.027 \\ 
\hline
\hline
\label{table1}
\end{tabular}
\end{table}

\begin{figure}[b]
\includegraphics[width=\linewidth]{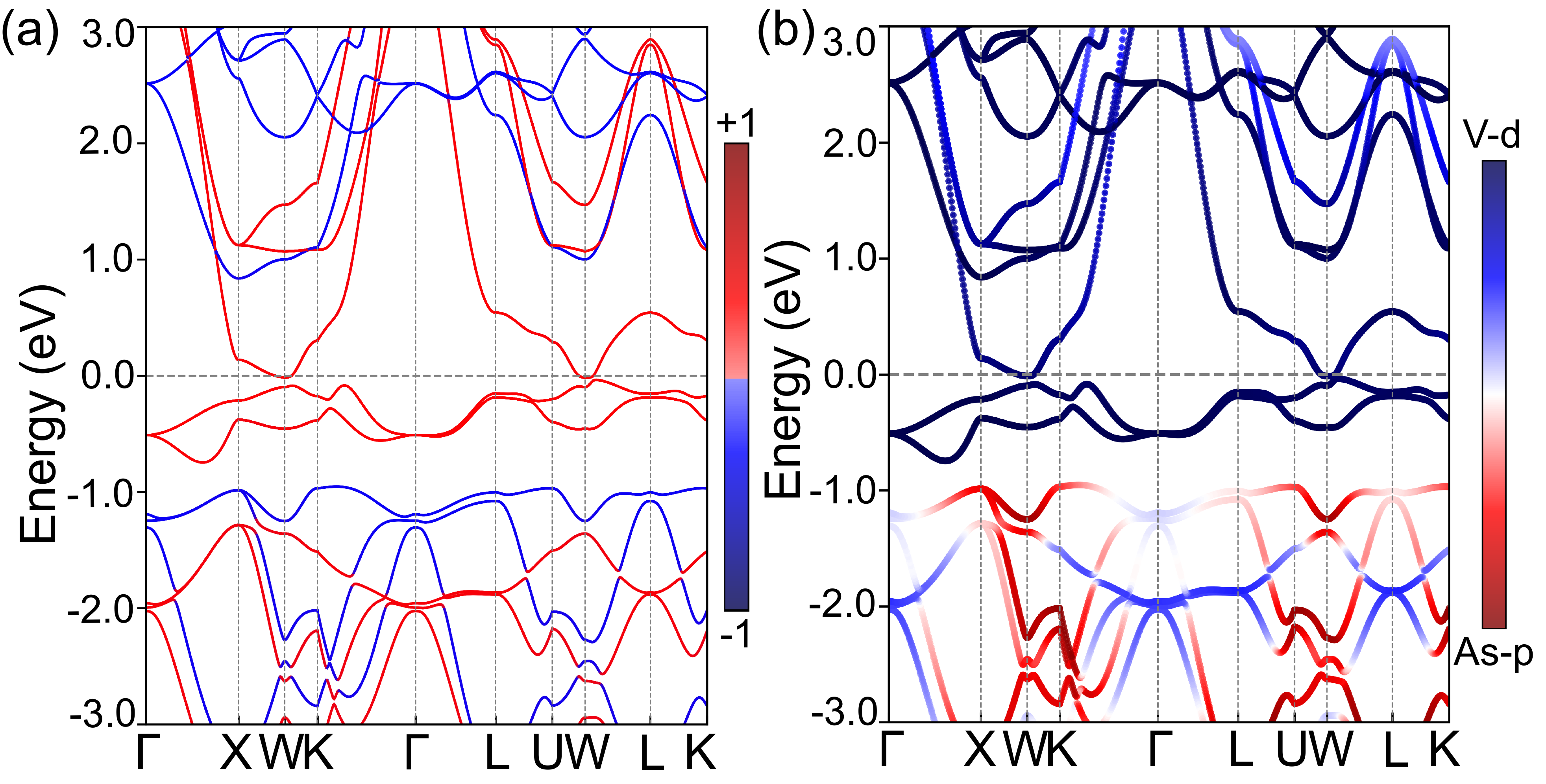}
\caption{(a) The spin projected band structures and (b) projected band structure of V-d orbitals and As-p orbitals of VAs using PBE functionals with considering SOC effect in the magnetization of [001].}
\label{fig4}
\end{figure}

\begin{figure*}[!t]
\includegraphics[width=\linewidth]{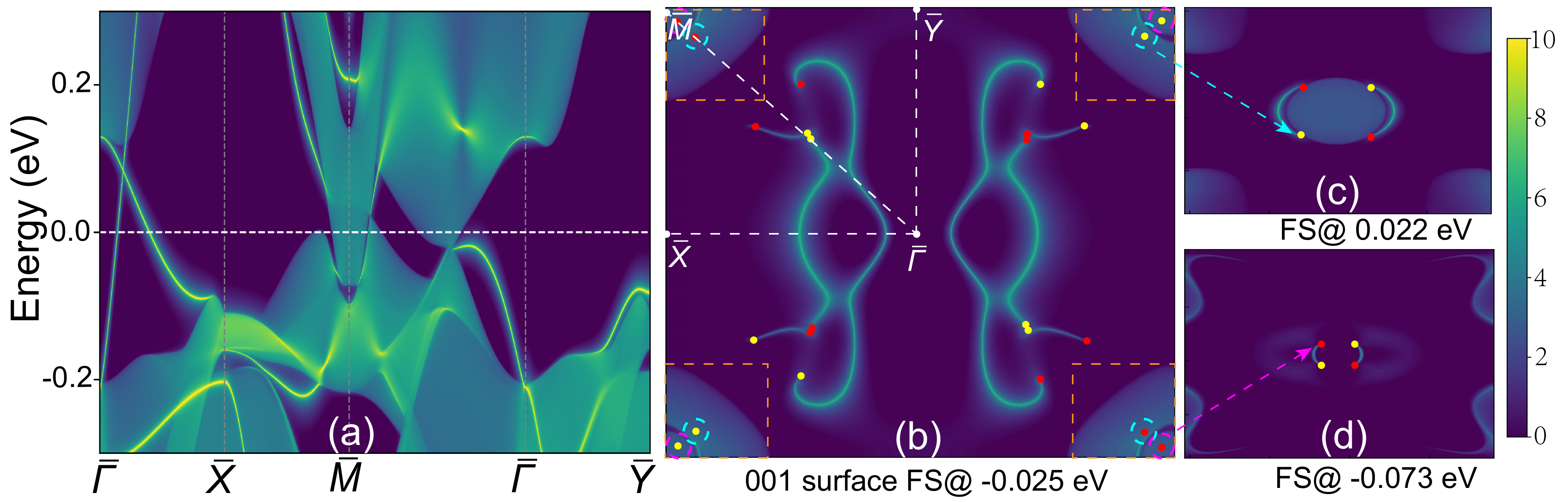}
\caption{Topological surface states and Fermi arcs projected on semi-infinite (001) and (111) surface of VAs. (a) The Local density of states (LDOS) for (001) surface of VAs in the magnetic direction [001]. The paths for LDOS in panels (a) are marked with white dashed lines in panels (b). (b-d) The projected Fermi surface for VAs in magnetic direction [001] at three different energies of -0.025 eV, 0.022 eV, and -0.073 eV, respectively. The red and yellow dots represent the projections of WPs with the topological charge of $\mathcal{C}$ = 1 and $\mathscr{C}$ = -1, respectively.}
\label{fig5}
\end{figure*}

The magnetic anisotropy energy differences in different magnetic directions of VAs compound are displayed in Fig. 3. Fig. \ref{fig3}(b) suggests that the low-energy magnetic states are in the axis directions. As displayed in Fig. 3(a), we fix $\phi$ in 45 $\degree$ and change $\theta$ from 0$\degree$ to 180$\degree$ and find that the lowest magnetic order is along the [001] direction. Thus, in the following sections, we will center on the magnetic direction of [001]. The band structures along high symmetry paths are plotted in Fig. \ref{fig3} in consideration of the SOC effect (The first BZ and the high symmetry paths are marked by red arrows in Fig. S2. Fig. \ref{fig3}(a) shows the spin-projected band structure, the red line represents the spin-up component and the blue lines represent the spin-down component. The spin-resolved band structure illustrates that the spin components are clearly separated because of the weak SOC interactions in VAs system. The electronic contributions near the fermi level are completely spin-up components, denoting the fully spin-polarized half-metallic features of VAs. Meanwhile, the projected electronic band structure of V-d and As-p orbitals of VAs are displayed in Fig. \ref{fig3}(b). It suggests that the main contribution around the fermi level is V-d orbitals.

\begin{figure}[!t]
\includegraphics[width=\linewidth]{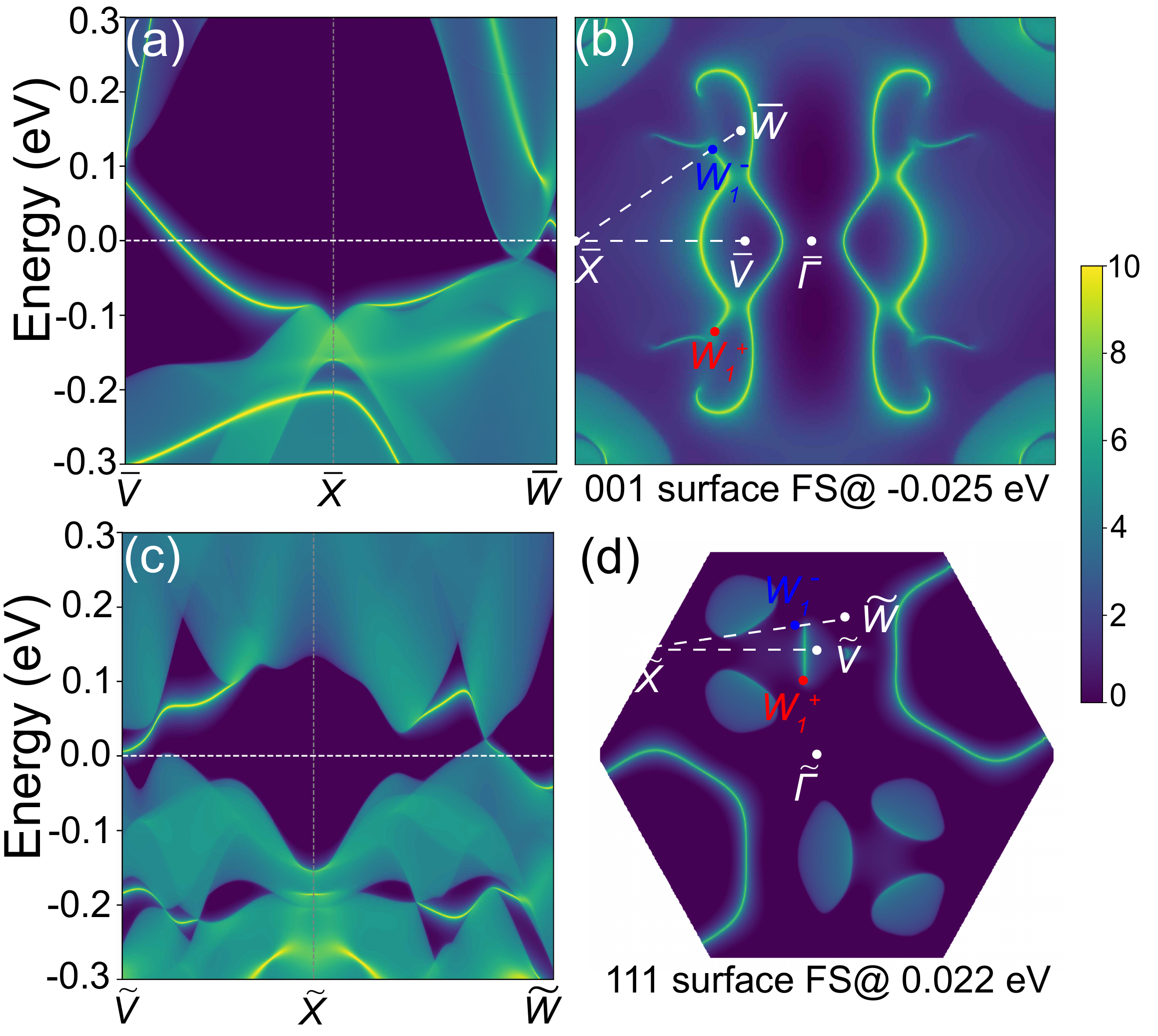}
\caption{Topological surface states and Fermi arcs projected on a semi-infinite (001) and (111) surface of VAs. (a, c) The topological surface states for (001) and (111) surface of VAs in the magnetic direction [001], in which the high symmetry paths for the corresponding surface are marked by white dashed lines in (b, d). (b, d) The projected Fermi surfaces for VAs, in which the energies are marked at the bottom of the corresponding figs.}
\label{fig6}
\end{figure}

In this section, we focus on investigating the topological semimetallic properties of VAs. Fig. S3(b) in the SM \cite{SM} are the comparisons of the band structures obtained from PBE functionals and the MLWFs in consideration of the SOC effect. The fitting precision of MLWFs is in good agreement with the PBE results, suggesting the validation of the following calculations. The surface states and Fermi arcs for both (001) and (111) surfaces (Fig. \ref{fig5} and Fig. \ref{fig6}) are obtained based on the Green function method \cite{PhysRevLett.119.036401, M_P_Lopez_Sancho_1984}. Fig. \ref{fig5}(a-d) are the projected surface states and fermi surfaces for the magnetic direction [001] of VAs. Fig. \ref{fig5}(a) shows the projected surface states on (001) surfaces, in which the reciprocal high-symmetry points and paths of the projected surface are marked by white points and dotted lines in Fig. \ref{fig5}(b). By carefully checking the energy differences between the lowest conduction and the highest valence bands, we find that there are twelve pairs of WPs over the entire BZ. The coordinates of all twelve pairs of WPs are obtained and tabulated in Table. \ref{table1}. The coordinates of these WPs are asymmetrically distributed in the first BZ, due to the lack of time-reversal and inversion symmetries. The energies of these WPs are very close to the fermi energy ie., -0.073, -0.025, and 0.022 eV, respectively. Fig. \ref{fig5}(b-d) shows the corresponding fermi surfaces projected on the (001) surface of VAs. The fermi arcs connected by opposite topological charges ($\pm{1}$) of WPs (red for $+1$ and yellow for $-1$) are clearly visible among eight pairs of them while the fermi arcs of the other four pairs of WPs (two pairs surrounded by cyan dotted circles and two pairs by yellow dotted circles) are covered by bulk states and can be visible in energies of 0.022 eV (Fig. \ref{fig5}(c)) and -0.073 eV (Fig. \ref{fig5}(d)), respectively. Meanwhile, the high energy magnetic order of magnetic direction [111] direction of VAs is also investigated and displayed in Fig. S4 of the SM \cite{SM}. As listed in Table. S1, there are still exist twelve pairs of WPs in the entire BZ and the energies and positions of these WPs are slightly different from the corresponding results in the magnetic direction [001] of VAs. These results reveal that the magnetic order can only move the WPs and does not eradicate these WPs, suggesting the robustness of these WPs in VAs system. The WPs shown in Fig. S4(e) along high symmetry lines of $\overline{M}$-$\overline{\Gamma}$ are even closer to the fermi energy than previous results in Fig. \ref{fig5}(a). Meanwhile, the Fermi arcs of all the WPs in magnetic direction [111] can be clearly observed in the energies of 0.0 eV and -0.05 eV as displayed in Fig. S4(b) and S4(c). 

Next, we select two energies (ie. -0.025 eV and 0.022 eV) with respect to the fermi energy to clearly see the topological surface states. In Fig. \ref{fig6}, Fig. \ref{fig6}(a-d) shows the results of the surface states and Fermi surfaces which are projected on (001) and (111) surfaces. The high symmetry paths calculated in Fig. \ref{fig6}(a) and Fig. \ref{fig6}(c) are marked by white dashed lines in Fig. \ref{fig6}(b) and Fig. \ref{fig6}(d). In (001) surface, the Dirac cone along $\overline{X}$-$\overline{\Gamma}$ is clearly visible, and the surface states connect the valance band and the conduction band. The projected Fermi arcs in Fig. \ref{fig6}(b) along $\overline{V}$-$\overline{X}$ can also be visible in Fig. \ref{fig6}(a). The projected surface states and Fermi arcs on the (111) surface in Fig. \ref{fig6}(c) and Fig. \ref{fig6}(d) also give clearly results. These results demonstrate that VAs is a Weyl semimetal with twelve pairs of WPs. Combined with the half-metal features of VAs, it can be concluded that VAs is a fully spin-polarized semimetal ferromagnet. As the energies of WPs are very close to the fermi level, it will be better for direct experimental observations.

In summary, based on DFT and DMFT theory, we investigate the half-metallic ferromagnet VAs which are confirmed to be a half-metallic ferromagnet with dynamic correlations. Considering the SOC effect, the magnetic anisotropic energy of VAs suggests the possible ground state magnetic orders are along [001] directions. The magnetic moment is obtained to 2 $\mu$B for the V element per unit formula. The topological surface states and Fermi arcs projected on semi-infinite (001) and (111) surfaces are investigated, and the result shows that the change of magnetic direction can only affect the positions of the Weyl points. There are twelve pairs of Weyl points with opposite topological charges of $+1$ and $-1$ over the entire BZ in both the magnetic direction [001] and [111]. The Fermi arcs of all the Weyl points can be observable in different energies and surfaces, confirming its magnetic topological Weyl semimetal properties. The fully spin-polarized weyl semimetal properties of zinc-blende compound VAs provide promising data for fundamental research in many-body physics and spintronic applications.

This work was supported by the National Natural Science Foundation of China (NSFC, Grants No. 12222402 and No. 92365101) and the Natural Science Foundation of Chongqing (Grant No. CSTB2023NSCQ-JQX0024). Simulations were performed on Hefei advanced computing center.
    
%

\end{document}


\title{Supplemental material for "Half-Metallic Ferromagnetic Weyl Fermions Related to Dynamic Correlations in the Zinc-blende Compound VAs"}

\author{Xianyong Ding}
\altaffiliation{X. Y. D. and X. J. contributed equally to this work.}
\affiliation{Institute for Structure and Function $\&$ Department of Physics, Chongqing University, Chongqing 400044, China}
\affiliation{Center for Quantum materials and devices, Chongqing University, Chongqing 400044, China}

\author{Xin Jin}
\altaffiliation{X. Y. D. and X. J. contributed equally to this work.}
\affiliation{College of Materials Science and Engineering, Chongqing Normal University, Chongqing 400044, China}

\author{Haoran Wei}
\affiliation{Institute for Structure and Function $\&$ Department of Physics, Chongqing University, Chongqing 400044, China}
\affiliation{Center for Quantum materials and devices, Chongqing University, Chongqing 400044, China}

\author{Ruixiang Zhu}
\affiliation{Institute for Structure and Function $\&$ Department of Physics, Chongqing University, Chongqing 400044, China}
\affiliation{Center for Quantum materials and devices, Chongqing University, Chongqing 400044, China}

\author{Xiaoliang Xiao}
\affiliation{Institute for Structure and Function $\&$ Department of Physics, Chongqing University, Chongqing 400044, China}
\affiliation{Center for Quantum materials and devices, Chongqing University, Chongqing 400044, China}

\author{Xiaozhi Wu}
\affiliation{Institute for Structure and Function $\&$ Department of Physics, Chongqing University, Chongqing 400044, China}
\affiliation{Center for Quantum materials and devices, Chongqing University, Chongqing 400044, China}

\author{Fangyang Zhan}
\email{zhan_fyang@cqu.edu.cn}
\affiliation{Institute for Structure and Function $\&$ Department of Physics, Chongqing University, Chongqing 400044, China}
\affiliation{Center for Quantum materials and devices, Chongqing University, Chongqing 400044, China}

\author{Rui Wang}
\email{rcwang@cqu.edu.cn}
\affiliation{Institute for Structure and Function $\&$ Department of Physics, Chongqing University, Chongqing 400044, China}
\affiliation{Center for Quantum materials and devices, Chongqing University, Chongqing 400044, China}

\maketitle
\section{Theoretical background of DFT + DMFT}
In order to investigate the dynamical correlation effect in VAs. We use DMFTwDFT package \cite{SINGH2021107778} based on the Dynamical Mean Field Theory (DMFT) which includes strong correlations beyond the static DFT exchange-correlation functional \cite{PhysRevB.69.245101, RevModPhys.78.865} to solve a single-site Anderson impurity problem embedded in an electronic bath ensured self-consistently as the original lattice is approximated to a quantum impurity problem \cite{PhysRevLett.69.168, shinaoka2017quantum}. The formula of DFT + DMFT functional $\Gamma$ can be written as:
\begin{equation}
\begin{array}{l}
\Gamma\left[\hat{\rho}, \hat{V}^{H x c}, \hat{G}^{c o r}, \hat{\Sigma}\right]  = -\operatorname{Tr}\left(\operatorname { l n } \left[\left(i \omega_{n}+\mu\right) \hat{1}- \hat{H}^{K S}\right.\right. \\
\left.\left.-\hat{P}_{\text {cor }}^{\dagger}\left(\hat{\Sigma}-\hat{V}^{D C}\right) \hat{P}_{c o r}\right]\right)+\Phi^{D F T}[\hat{\rho}]-\operatorname{Tr}\left(\hat{V}^{H x c} \hat{\rho}\right) \\
+\Phi\left[\hat{G}^{c o r}\right]-\operatorname{Tr}\left(\hat{\Sigma} \hat{G}^{c o r}\right)-E^{D C}\left[\hat{G}^{c o r}\right]+ \operatorname{Tr}\left(\hat{V}^{D C} \hat{G}^{c o r}\right)
\end{array}
\end{equation}
where $\omega_{n}$ is the Matsubara frequency for fermions, $\mu$ is the chemical potential, $\phi^{DFT}[\rho]$ is the DFT interaction energy, $\hat{H}^{KS}$ is the $Kohn$-$Sham$ $(KS)$ Hamiltonian operator, $\hat{V}^{Hxc}$ is the $Hartree$-$exchange$-$correlation$ $(Hxc)$ potential operator, $\hat{V}^{DC}$ is the double-counting (DC) potential operator, $\hat{P}_{cor}$ ($\hat{P}_{cor}^{\dagger}$) is a projection operator defined to downfold (upfold) between the correlated subspace which is composed of correlated orbitals and the hybridization subspace which hybridize both correlated orbitals and uncorrelated orbitals. The DMFT interaction energy, $\phi[G^{cor}]$ is the luttinger-Ward functional summing all vacuum-to-vacuum Feynman diagrams which are local \cite{PhysRev.118.41, PhysRev.118.1417, PhysRev.121.942}. $E^{DC}$ is the double-counting (DC) term. In this work, the expression of the DC term can be written as:
\begin{equation}
E^{DC} = \frac{(U-\alpha)}{2} \cdot N_{d} \cdot (N_{d} - 1) - \frac{J}{4} \cdot N_{d} \cdot (N_{d} - 2)
\end{equation}
where $U$ is the on-site Hubbard interaction, J is the Hund`s coupling, and $N_{d}$ is the occupancy of correlated orbitals within the correlation subspace. The elaborate computational methods can be found in literature \cite{SINGH2021107778}.

\begin{figure}[!htb]
\includegraphics[width=\linewidth]{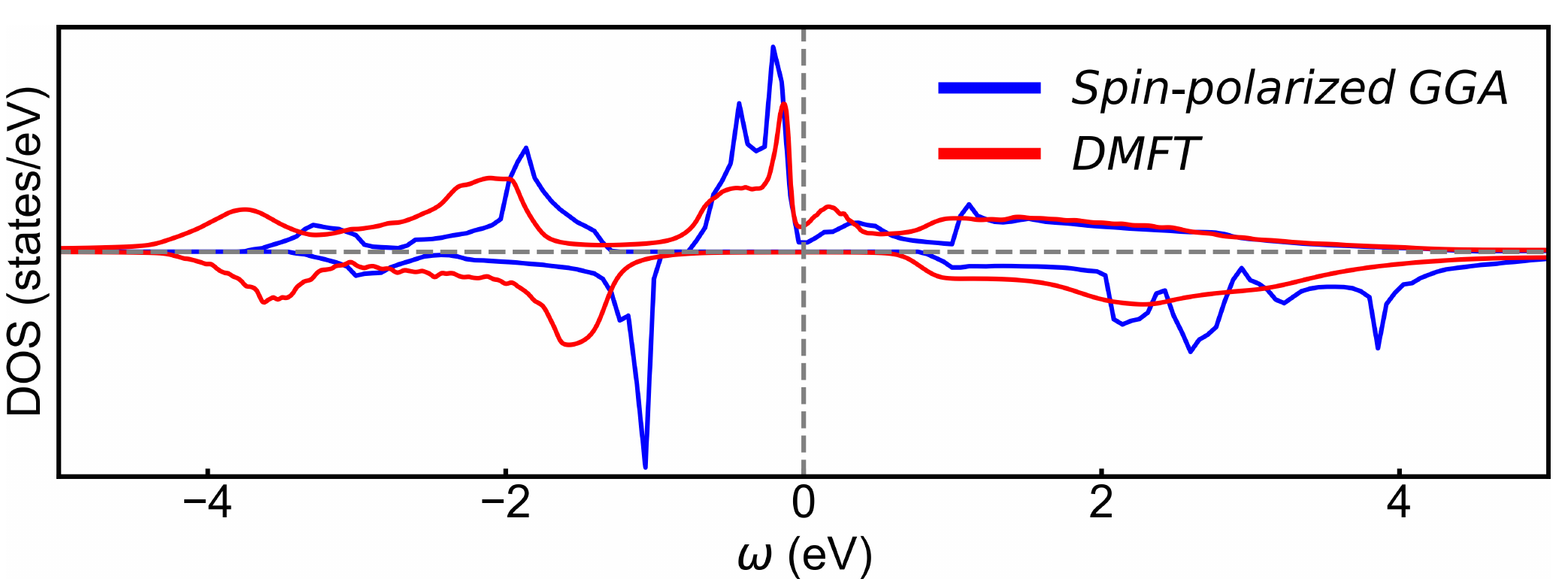}
\caption{The summed DOS of V-d orbitals and As-p orbitals are obtained from DFT + DMFT and DFT + $U$ methods without considering the SOC effect.}
\label{figS1}
\end{figure}

\section{Computational Details}
Vienna Ab initio Simulation Package (VASP) is used for implementing electronic structure calculations using the projector augmented wave (PAW) method  \cite{PhysRevB.54.11169}. The generalized gradient approximation with Perdew-Burke-Ernzerhof (PBE) functional is used to describe the exchange-correlation interaction \cite{KRESSE199615, PhysRevLett.77.3865}. The cutoff energy for plane wave functions is set to 500 eV. The first Brillouin zone (BZ) was sampled by $11\times 11 \times 11$ $\Gamma$-centered k grid for VAs. The convergence for the forces of all atoms are relaxed to be less than $1.0\times10^{-3}$ eV/{\AA}. DFT + DMFT method are performed in a DMFTwDFT package \cite{SINGH2021107778} in order to consider the dynamical correlation effect between $e_{g}$ and $t_{2g}$ of V-d orbitals. DFT + $U$ method is performed with effective $U$ values of 1.5 eV for V atoms to calculate the electronic properties of VAs. The spin-orbital interactions are also considered in the same time. Moreover, to reveal topological electronic features of VAs, a tight-binding (TB) Hamiltonian is constructed by projecting the Bloch states into Maximally Localized Wannier Functions (MLWFs) in the WANNIER90 package \cite{RevModPhys.84.1419, MOSTOFI20142309}.

Fig. \ref{figS1}(a) gives the summed density of states of V-d orbitals and As-p orbitals by using DFT + U (blue lines) method and DFT + DMFT method (red lines). The result shows that around the energy of [-1, 1], the trend of the density of states between DFT + $U$ and DFT + DMFT is basically consistent. Furthermore, the results of DFT + DMFT are also consistent with previous results \cite{SINGH2021107778}.

\begin{table}[hb] 
\caption{\label{weylpoints} The corresponding positions, topological charges, and energies of the WPs in the magnetization direction [111] of VAs.}
\renewcommand\tabcolsep{14 pt}
\begin{tabular}{ccr}      
\hline
\hline
     Coordinate (x, y, z)  & Charge   & E-E$_{F}$ (eV) \\
\hline
        (0.47, -1, 0.0)      &  -1    & 0.007 \\ 
        (-0.46, 0.0, -1)      &   1    & -0.001 \\ 
        (0.5, 0.0, 1.0)      &   1    & -0.05 \\ 
        (-0.5, -1.0, 0.0)      &  -1    & -0.05 \\ 
        (-0.5, 1.0, 0.0)      &  -1    & -0.054 \\ 
        (0.0, -1.0, -0.46)      &   1    & -0.001 \\ 
        (0.0, -0.5, -1.0)      &  -1    & -0.05 \\ 
        (0.5, 0.0, -1.0)      &   1    & -0.054 \\ 
        (-1.0, 0.5, 0.0)      &   1    & -0.054 \\ 
        (0.46, 1.0, 0.0)      &  -1    & -0.001 \\ 
        (-0.47, 0.0, 1.0)      &   1    & 0.007 \\ 
        (0.0, -1.0, 0.5)      &   1    & -0.054 \\ 
        (1.0, 0.0, -0.5)      &  -1    & -0.054 \\ 
        (0.0, 1.0, 0.5)      &   1    & -0.05 \\ 
        (0.0, 0.6, 1.0)      &  -1    & -0.001 \\ 
        (0.0, 0.47, -1.0)      &  -1    & 0.007 \\ 
        (1.0, -0.47, 0.0)      &   1    & 0.007 \\ 
        (1.0, 0.5, 0.0)      &   1    & -0.05 \\ 
        (-1.0, 0.0, -0.5)      &  -1    & -0.05 \\ 
        (-1.0, -0.46, 0.0)      &   1    & -0.001 \\ 
        (1.0, 0.0, 0.46)      &  -1    & -0.001 \\ 
        (-1.0, 0.0, 0.47)      &  -1    & 0.007 \\ 
        (0.0, -0.5, 1.0)      &  -1    & -0.054 \\ 
        (0.0, 1.0, -0.47)      &   1    & 0.007 \\ 

\hline
\hline
\label{tableS1}
\end{tabular}
\end{table}

\begin{figure}[!htb]
\includegraphics[width=0.6\linewidth]{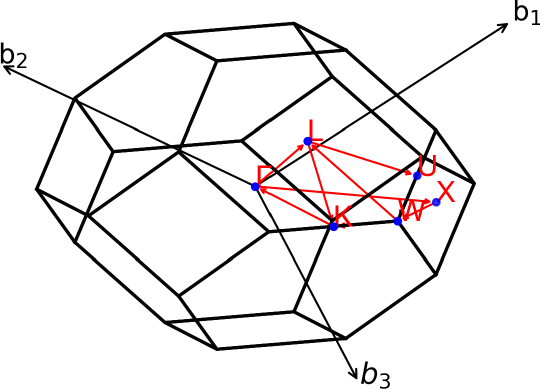}
\caption{The first BZ of VAs and red words mark the high symmetry points.}
\label{figS2}
\end{figure}

\begin{figure}[!htb]
\includegraphics[width=\linewidth]{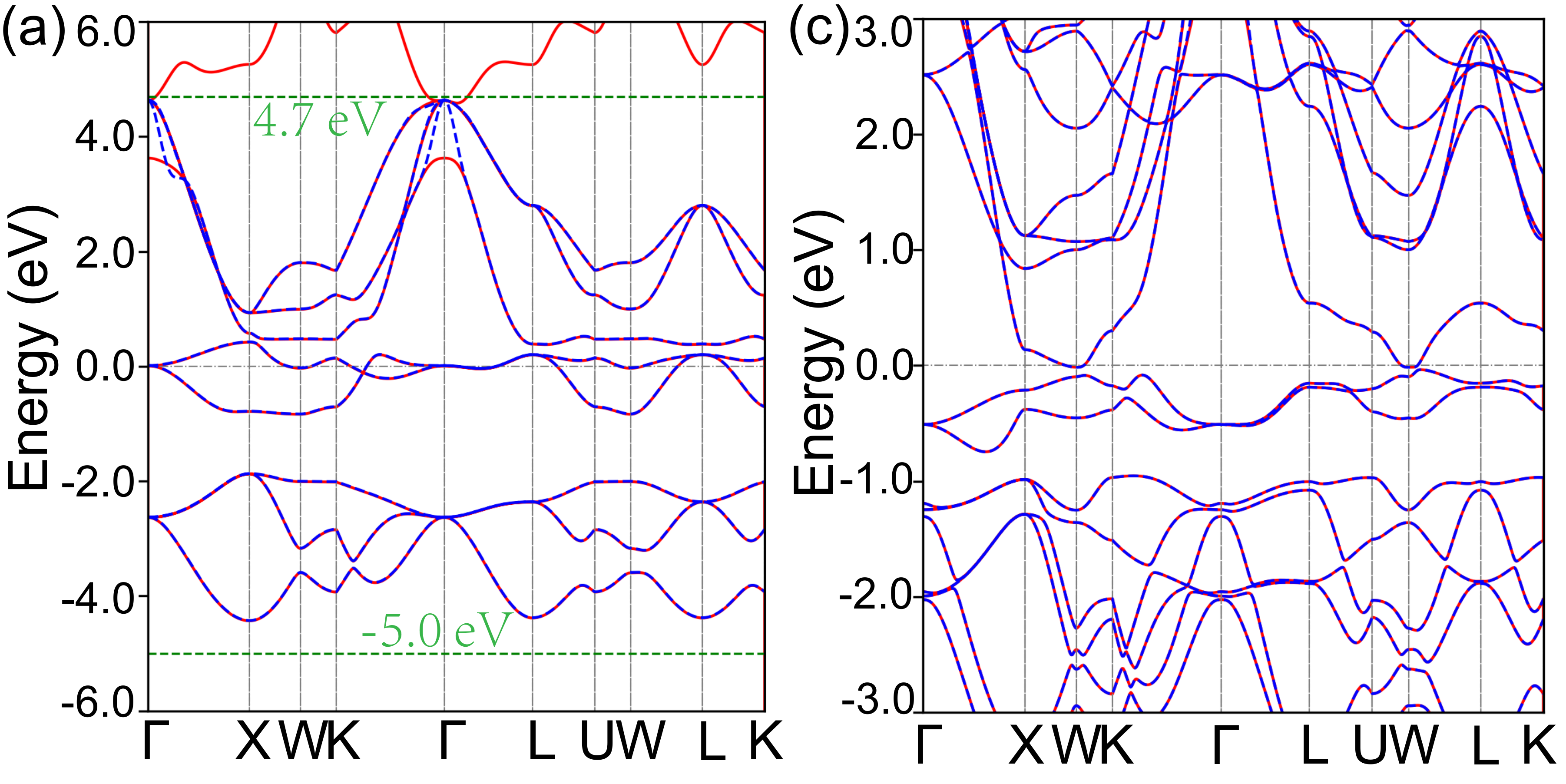}
\caption{(a) The comparison of the band structures of VAs by using PBE functionals and Wannier TB Hamiltonian in which (a) without considering spin-polarization, and (b) the magnetic direction [001] of VAs in the consideration of SOC effect.}
\label{figS3}
\end{figure}

Fig. \ref{figS3}(a) shows the comparison between the results from PBE functionals and Wannier Tight-binding (TB) Hamiltonian, in which Fig. \ref{figS3}(a) do not consider spin polarization. Fig. \ref{figS3}(b) is the results in the magnetic direction [001] of VAs in consideration of the SOC effect. Obviously, the results of the Wannier TB Hamiltonian are completely consistent with the DFT + $U$ results, confirming the reliability of the computed Wannier functions.

\begin{figure*}[!t]
\includegraphics[width=\linewidth]{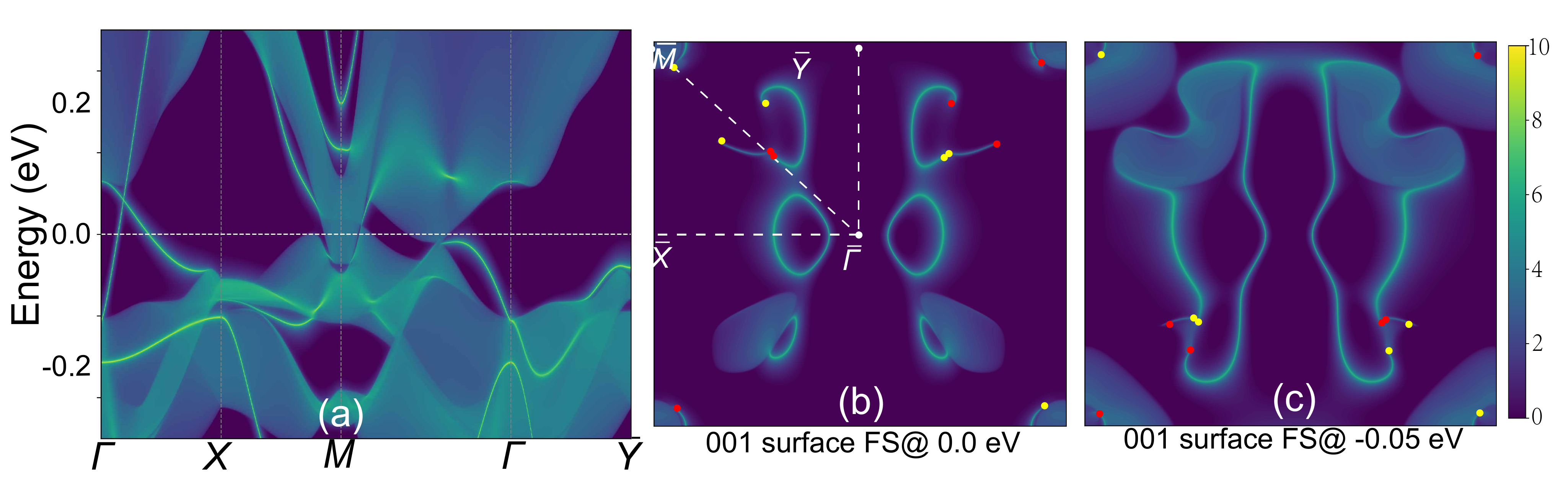}
\caption{Topological surface states and Fermi arcs projected on semi-infinite (001) and (111) surface of VAs. (a) The Local density of states (LDOS) for (001) surface of VAs [111]. The paths for LDOS in panels (a) are marked with white dashed lines in panels (b). (b, c) The projected Fermi surface for VAs in magnetic direction [111] at three different energies of -0.025 eV, 0.022 eV, and -0.073 eV, respectively. The red and yellow dots represent the projections of WPs with the topological charge of $\mathcal{C}$ = 1 and $\mathscr{C}$ = -1, respectively.}
\label{figS4}
\end{figure*}

\newpage